\documentclass[a4paper,12pt]{article}
\usepackage{latexsym}
\usepackage{graphicx}
\usepackage{dcolumn}
\usepackage{bm}
\usepackage[T1]{fontenc}
\usepackage[ansinew]{inputenc}
\usepackage{amsmath}
\usepackage{amssymb}
\usepackage{graphicx}
\begin{document}
\title{`Deterministic' quantum plasmonics}
\date{\today}
 \maketitle
Author Names: \emph{Aur\'{e}lien Cuche, Oriane Mollet, Aur\'{e}lien Drezet, and Serge Huant}\\
Author address: \emph{Institut N\'eel, CNRS and Universit\'e Joseph Fourier Grenoble, BP 166, 38042 Grenoble Cedex 9, France}\\
\\
Contacts:\\ aurelien.cuche@grenoble.cnrs.fr\\oriane.mollet@grenoble.cnrs.fr\\aurelien.drezet@grenoble.cnrs.fr\\serge.huant@grenoble.cnrs.fr\\
Abstract:\\

We demonstrate `deterministic' launching of propagative quantum surface-plasmon polaritons at freely chosen positions on gold plasmonic receptacles. This is achieved by using as plasmon launcher a near-field scanning optical source made of a diamond nanocrystal with two Nitrogen-Vacancy color-center occupancy. Our demonstration relies on leakage-radiation microscopy of a thin homogeneous gold film and on near-field optical microscopy of a nanostructured thick gold film. Our work paves the way to future fundamental studies and applications in quantum plasmonics that require an accurate positioning of single-plasmon sources and may open a new branch in plasmonics and nanophotonics, namely scanning quantum plasmonics.\\


\newpage

Quantum plasmonics - quantum optics with surface-plasmon polaritons (SPPs)~\cite{Barnes,Novotny} - is an emerging field with valuable prospects both on the fundamental science and application agenda. One topic of large impact is concerned with the coupling of single quantum emitters such as molecules, quantum dots, or nanodiamonds with plasmonic devices. In this context, early and more recent studies have shown that fluorescent quantum emitters can efficiently couple to SPPs when they are located in the vicinity of a metal structure~\cite{Drexhage,Drexhage2,Anger,Bhara,Gerber}. Moreover, the possibility to generate individual SPPs with single-photon sources opens the door to a wide range of studies such as single-SPP mediated energy transfer~\cite{Chang,Akimov,Fedutik,Wei}, locally-controlled enhanced fluorescence~\cite{Chang,Schie}, or single-SPP interferometry~\cite{Kolesov}. Still, a fundamental understanding together with a tight control in space, energy and polarization within this quantum regime is essential to fully exploit these stimulating promises. Ideally, this requires a deterministic control on the coupling of selected quantum emitters to tailored plasmonic structures~\cite{Chang,GerardCdF,Liu}. Here, we make a decisive step forward in this direction by demonstrating deterministic launching of propagative quantum-SPPs at well-defined and freely chosen positions into a nano-structured metal film. \\
Our method contrasts with previous related works~\cite{Akimov,Fedutik,Wei,Schie,Kolesov} which all used randomly dispersed quantum emitters to couple with localized plasmonic modes confined into nanowires or particles. \begin{figure}[h]
\begin{center}
\begin{tabular}{c}
\includegraphics[width=8cm]{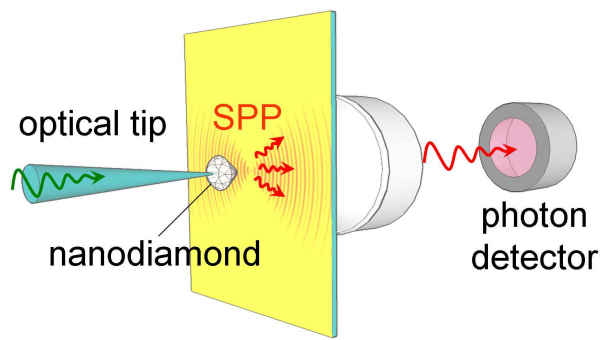}
\end{tabular}
\end{center}
\caption{Principle of near-field scanning quantum plasmonics. A laser light impinging trough an optical fiber tip excites the fluorescence of (two) NV color-centers hosted by a nanodiamond that is glued at the apex of an optical tip. Single fluorescence photons converted into single SPPs propagate along a gold air interface. SPPs are subsequently converted back into photons and are collected by a microscope objective before being sent to a photon detector.}
\end{figure}It relies on the use of (two) quantum emitters fixed in a well-controlled way onto the tip apex of a near-field scanning optical microscope (NSOM)~\cite{Cuche,Cuche2} as sketched in Figure~1. This work takes benefit from recent developments on ``active-tip based NSOM''~\cite{Mich,Che} (see~\cite{Novotny} for a review on NSOM). In particular, a scanning quantum source made of a single-photon emitter, i.e. a single Nitrogen-Vacancy (NV) color-center hosted by a 20 nm diamond crystal that is grafted onto a NSOM tip apex, has recently been introduced~\cite{Cuche}. Thanks to the extraordinary photo-stability of this quantum emitter at room temperature~\cite{Cuche,Brouri,Yannick}, the NV-based NSOM probe builds an ideal platform for the on-demand local launching of single SPPs as reported here.\\
We first demonstrate SPP launching along the air-metal interface of an unstructured thin 60 nm gold film by using the so-called leakage radiation microscopy (LRM) mode~\cite{Hecht,Bouhelier,Drezet2} (see Figure~1). The physical basis of our approach can be understood by recalling that SPPs at the air-metal interface are characterized by a complex in-plane wavevector
\begin{eqnarray}
K_{\textrm{SPP}}(\lambda)=\frac{2\pi}{\lambda}\sqrt{\epsilon(\lambda)/(\epsilon(\lambda)+1)}
\end{eqnarray} that depends on the optical wavelength $\lambda$ and on the metal dielectric permittivity $\epsilon(\lambda)$. Due to boundary conditions and conservation of the in-plane wavevector along the different interfaces, SPPs leak through the thin gold film into the glass substrate at a fixed angle $\Theta_{LR}$ with $n\sin{(\Theta_{LR})}\simeq \textrm{Real}[K_{\textrm{SPP}}(\lambda)]\lambda/(2\pi)$ and $n\simeq 1.5$ the glass optical index. In the experiments these leaky waves are detected using an immersion oil objective ($100\times$, NA=1.4) located below the sample and are directly imaged with a Peltier cooled charged coupled device (CCD) camera.\\
This LRM detection mode~\cite{Hecht} constitutes a direct method for mapping the SPP propagation and thus for monitoring the tip ability to efficiently launch SPPs. Noteworthy, SPPs being emitted at a fixed angle $\Theta_{LR}$ their intensity profile spans a circle in the transverse Fourier or momentum space. To access this information, we include in the LRM detection path a set of lenses for imaging either the Fourier plane (FP) or the direct propagation of SPPs in the image plane (IP)~\cite{Hecht,Bouhelier,Drezet2} (Supporting Information).\\
As a test of the proposed experimental scheme we consider a standard aluminum coated NSOM tip with a 100 nm clear aperture at the apex~\cite{Novotny}. Such a classical tip can actually act as a localized SPP source when it is brought into close proximity to a metal film~\cite{Hecht,Sonnichsen,Brun}, i.e. at 20-30 nm above the surface. Figures~2 a and 2 b show \begin{figure}[h]
\begin{center}
\begin{tabular}{c}
\includegraphics[width=8cm]{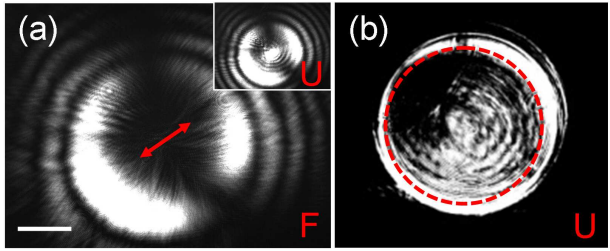}
\end{tabular}
\end{center}
\caption{(\textbf{a}) and (\textbf{b}) LRM images in the IP and FP, respectively, for an aperture NSOM tip launching SPPs along a 60 nm thick gold film. U and F stand for the Fourier unfiltered and filtered cases, respectively, i.e. images are acquired without and with a beam-blocker positioned in the IP, respectively. The size of the beam blocker is illustrated by a red dotted circle in (\textbf{b}). The scale bar in (\textbf{a}) is 10$\mu$m long. The fringes in (\textbf{a}) are due to intrinsic limitations of our commercial microscope and explain that Fourier filtering the 'allowed' light in the F image removes the leakage radiation from the position of the tip, thereby artificially creating an intensity null at the center.}
\end{figure}respectively the LRM IP and FP images recorded with the NSOM tip immersed in the near-field of the metal film (the $\lambda=647$ nm line of an Argon-Krypton laser is used). As seen in the IP image SPPs propagate away from the tip with an estimated damping constant $L_{\textrm{SPP}}=1/(2\textrm{Imag}[K_{\textrm{SPP}}])\simeq 10-20$ $\mu$m (this value is deduced from fitting an intensity crosscut taken in Figure 2 a with the SPP point source intensity profile~\cite{Drezet2}) and draw a circle of radius  $\delta k= \textrm{Real}[K_{\textrm{SPP}}]$ in the FP. The two SPP lobes imaged in both the IP and FP are associated with the dipolar nature of the NSOM tip~\cite{Hecht,Brun}. The interference fringes seen in the IP are residual intrinsic artifacts attributed to our commercial optical microscope setup~\cite{Stepanov}. It is worth pointing out that the light observed inside the SPP circle in Figure~2 b is the allowed light emitted through the sample at smaller angles than the total internal reflection angle in glass $\Theta_c=\arcsin{(1/n)}\simeq 42^{\circ}$. Since $\Theta_{LR}>\Theta_c$ this allows us to implement a Fourier filter~\cite{Drezet2} to selectively block the allowed light in the Fourier plane. The size of the circular beam blocker is shown in Figure~2 b. The effect of Fourier filtering (labeled F) in the IP plane is already included in Figure~2 a and compared with the unfiltered configuration (labelled U) shown in the inset.\\
\begin{figure}[h]
\begin{center}
\begin{tabular}{c}
\includegraphics[width=8cm]{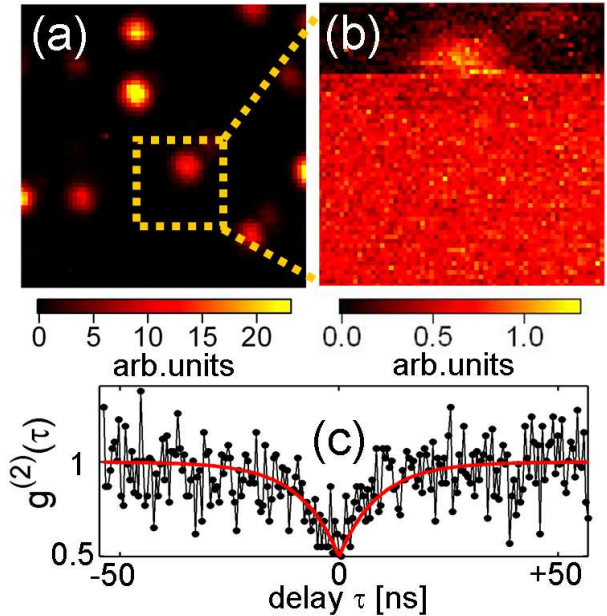}
\end{tabular}
\end{center}
\caption{(\textbf{a}) $5\times5$ $\mu$m$^2$ confocal fluorescence image of nanodiamonds dispersed on a glass substrate. (\textbf{b}) NSOM image of the diamond seen in (\textbf{a}) showing its capture by the scanning tip. (\textbf{c}) $g^{2}(\tau)$ correlation curve for the diamond trapped in (\textbf{b}) corresponding to 2 NV color centers. Such a double NV occupancy has been encountered with most of the nanodiamonds imaged in this work. The fit (red curve) is made using eq~2 and fulfils the purely quantum constraint $1/2\leq g^{2}(\tau)<1$.  }
\end{figure}
We now turn on to the SPP launching by a NV-based NSOM tip. Our
ambition of going to the quantum plasmonics realm within the
framework developed in this letter imposes using a tip with a
reduced number of NVs for SPP launching. Therefore, we consider a
tip with a \emph{single fluorescent nanodiamond} which, as it
turns out, hosts two NV centers (a vast majority of diamonds probed in the present study hosted two NV centers). Such a tip is
produced following the procedure described in
\cite{Cuche}. Briefly, a polymer-coated dielectric NSOM
probe traps during scanning a single 20 nm nanodiamond (Figure~3 b)
that has been first selected on a fluorescence confocal image
(Figure~3 a). The graft event translates into a persistent increase
of the fluorescence signal embarked by the scanning
tip~\cite{Cuche} (see Figure~3 b).\\
A stringent test of the ability of this nanodiamond-based tip to enter the quantum optics arena is provided
by the second-order time-intensity correlation function
$g^{(2)}(\tau)$ \cite{Cuche,Brouri,Yannick}. For a nanodiamond
hosting $N$ color centers, it reads
\begin{eqnarray}
g^{(2)}_N(\tau)=1-\frac{1}{N}e^{-(\Gamma+2\Gamma')\tau}
\end{eqnarray}
where $\Gamma$ and $\Gamma'$ are the spontaneous emission and pumping rates, respectively. The important point here is that $g^{(2)}_N\in [1-1/N,1]$ always takes numerical values  smaller than the classical bound $g_{\textrm{bound}}^{(2)}=1$. This photon anti-bunching phenomenon definitely demonstrates the quantum nature of the photon emission. Figure~3 c  depicts  $g^{2}(\tau)$ for the single diamond tip described above. It perfectly fits with eq~2 using $N=2$ and a lifetime $\Gamma^{-1}$ of 9 ns, in excellent agreement with earlier reports~\cite{Cuche}. This unambiguously shows that the functionalized tip hosts 2 NV centers exactly.\\
The quantum photon tip considered so far is subsequently used to launch SPPs into the thin gold film, the optical excitation of the NVs being ensured by shining a laser light (wavelength $\lambda_E=488$ nm) directly into the optical fiber that is terminated by the tip (Figure~1). The inclusion in the optical path of a band-pass filter transmitting light in the $\lambda\in[572\textrm{ nm}, 642\textrm{ nm}]$ range ensures that the only signal imaged by the CCD camera is initiated by the NV fluorescence. Note that at $\lambda_E=$ 488 nm SPPs cannot be efficiently launched into gold due to the strong absorption in the interband region~\cite{Novotny} which reduces the SPP propagation length below $L_\textrm{SPP}\simeq 100$ nm. Therefore, SPPs, if any, are launched by the NV fluorescence which has the appropriate properties (NVs emit in the red-orange part of the spectrum). The IP and FP images in Figures~4 a  and 4 b  are reminiscent of those in Figures~2 a  and 2 b. Thus, they demonstrate that SPPs are actually launched by the quantum photon source.\\
There are however phenomenological differences with the diamond source that are worth commenting on. First, there are no clear polarization lobes compared to the aperture tip. We assume this is likely due to the \begin{figure}[h]
\begin{center}
\begin{tabular}{c}
\includegraphics[width=8cm]{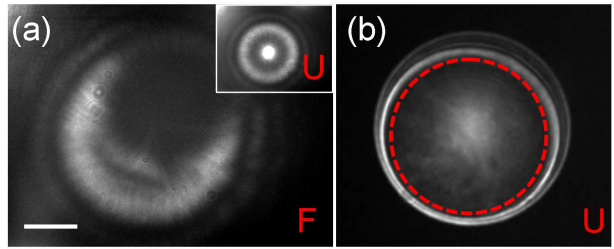}
\end{tabular}
\end{center}
\caption{(\textbf{a}) and (\textbf{b}) LRM images in the IP and FP, respectively, for the 2 color-center NSOM tip launching SPPs along the gold film. U, F and the red circle have the same meaning as in Figure~2. The scale bar in (\textbf{a}) is 10$\mu$m long. The partial vigneting of the tip emission that is visible in the right upper corner of (\textbf{a}) results from a residual misalignment of the Fourier filter. }
\end{figure}fact that the transition dipoles associated with the 2 NV centers have in this particular tip important vertical components that generate an isotropic angular SPP profile along the metal interface. Second, despite the rather large NV spectrum~\cite{Cuche,Brouri,Yannick}, the SPPs draw a well-defined circle in Fourier space. This circle is so bright that the allowed light is barely visible in the unfiltered Fourier image of Figure~4 b  compared with Figure~2 b. This is because in the relevant wavelength range, all LRM angles $\Theta_{LR}$ fall just above $\Theta_c$ within a reduced angular spreading $2^{\circ}$ wide only.\\ Before proceeding further, it is worth stressing that SPP launching by both the aperture and single diamond tips takes place in the near-field only. Keeping both tips at distances $\gtrsim 100$ nm above the gold film does not allow to reveal the typical SPP signatures of Figures~2 and 4. In this case, only the allowed light such as the far-field nanodiamond fluorescence can be imaged.\\
From a fundamental point of view the very fact that the LRM images obtained with the quantum source agree qualitatively well with those generated by the classical SPP source of Figure~2 is of significance since it confirms the wave nature of the emitted SPPs. We point out however that since $g^{(2)}(0)=1/2$ the launching source cannot emit more than two photons, and consequently
launch more than two SPPs at once (see Figure~3 c). This is a particle-like behavior which cannot be understood in the context of classical optics. Hence, the images of Figure~4 are a manifestation of the wave-particle duality for SPPs~\cite{Kolesov}. This entails that the quantum plasmonics realm has indeed been entered.\\
The above LRM study shows that single SPPs can be launched onto a flat and unstructured gold film by the near-field fluorescence generated by the NV-based tip. We now demonstrate that we actually have a `deterministic' control on the SPP launching position. For this purpose we turn on to a thick, i.e. opaque to visible light to avoid any direct light transmission, nanostructured film. This nanostructure (insert of Figure 5 a) is a ring-like aperture slit (1.8  $\mu$m inner diameter, 150 nm rim width) obtained by focused-ion-beam milling of a 200 nm gold film. The imaging tip is a 2-NV-center tip, different from that of Figure 4. Since the gold film is thick, LRM is no longer possible. Therefore we switch to the transmission-NSOM imaging mode using a $60\times, NA=0.95$ dry microscope objective (Supporting Information). The photon flux in the IP is collected by means of an optical fiber which is optically conjugated with the NSOM tip and defines a confocal collection region on the gold film. The collected signal is sent to an avalanche photodiode as explained in \cite{Cuche}. To interpret the NSOM images, it is worth remembering that like in any confocal microscope the imaged zone on the sample has an extension that is set by the core diameter of the collection fiber divided by the objective magnification ratio. Therefore, we will below make use of two fibers with very different core diameters in order to discriminate between the short-range near-field and long-range plasmon-field contributions.\\
Figure~5 a shows the NSOM  (optically unfiltered) image of the total power radiated by the
quantum source. The image is dominated by the excitation laser light at
$\lambda_E=$488 nm. It exhibits a
distinctive two-lobe structure which is reminiscent of the
incident light polarization at the tip apex. Figure~5 b shows a NSOM
filtered image
\begin{figure}[h]
\begin{center}
\begin{tabular}{c}
\includegraphics[width=8cm]{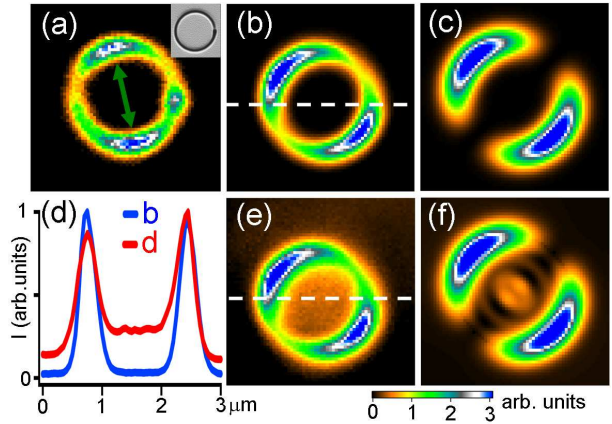}
\end{tabular}
\end{center}
\caption{(\textbf{a}) $3\times3$ $\mu$m$^2$ spectrally unfiltered NSOM image obtained by scanning the ring aperture slit seen in the scanning electron micrograph of the inset under the 2-NV-based tip. This image is dominated by the excitation laser light at 488 nm. The green arrow indicates the
incident laser polarization at the tip apex. (\textbf{b}) and (\textbf{e}) are the same as (\textbf{a}) but acquired in the NV fluorescence band after filtering of the incident laser light. They are obtained with 50 $\mu$m and 360
$\mu$m diameter collection fibers placed in the IP, respectively. (\textbf{d}) Normalized intensity crosscuts along the dashed dotted white
lines shown in (\textbf{b},\textbf{e}). (\textbf{c}) and (\textbf{f}) are numerical simulations of (\textbf{b}) and (\textbf{e}), respectively. }
\end{figure} in the
$\lambda\in[572\textrm{ nm}, 642\textrm{ nm}]$ range overlapping with the NV emission. In contrast with the LRM images in Figure~4, this image also clearly resolves a two-lobe structure which is however not aligned with that produced by the excitation and instead depends on the orientation of the NV-center transition dipoles. This suggests that the NV centers have
here strong in-plane components of their transition dipoles that
are not aligned with the excitation polarization (Note that the precise orientation of the transition dipoles cannot be controlled within our ND attachment procedure). This is further confirmed by a simulation (Figure~5 c) that neglects SPPs (see below) but takes into account the dipolar emission of the NVs scanning above the nanostructure. This model (Supporting Information) shows that only the in-plane components of elementary dipoles can generate a two-lobe pattern (compare Figures~5 b  and 5 c).\\
Importantly, the
diameter of the collection optical fiber in Figure~5 b is chosen so
that only a small area of $r=400$ nm radius around the tip
contributes to the optical signal during scanning of the nanostructure under the fixed tip~\cite{Cuche} (more details are given in the Supporting Information file).
This configuration precludes the observation of SPPs occupying the interior of the
planar cavity with a diameter much larger than $r$, which justifies the assumption made for computing Figure~5 c. However, shifting to a larger collection fiber
corresponding to a $r'=3.0$ $\mu$m radius area around the tip as
done in Figure~5 e  reveals a very different behavior. Since $r'$ is
larger than the ring cavity diameter, SPPs confined
inside the circular structure are now plainly visible as seen from
the intensity contrast in Figure~5 e (compare also the
intensity cross cuts in Figure~5 d). In
addition, the image is also blunt outside the ring area.  The
interpretation here is that SPPs are launched in a ``deterministic'' way, i.e. at the tip position, into the nanostructured gold film
by the NV-center quantum source and propagate towards the rim zone
where they are diffracted back to a photon signal that is collected on the other side of the thick opaque film to form a
transmission image. Therefore, the image of Figure~5 e includes the searched-for
quantum SPP information which is not accessible in Figure~5 b  and
gives evidence for a ``deterministic'' launching of SPPs by the quantum NV-based tip.\\
The main experimental features are reproduced in a simulation that extends the one discussed in Figure~5 c  by including the SPPs launched by the dipolar source during scanning (see Figure~5 f). In particular the SPP field confinement in the cavity is reproduced irrespective of the dipole orientations. Here, to reproduce qualitatively Figure~5 e, we have considered two dipoles emitting incoherently and having both in-plane and vertical components (Supporting Information).\\
In summary, we have shown in this letter that a photostable
quantum NSOM probe made of two NV centers hosted by a nanodiamond
is able to launch in a ``deterministic way'', i.e. at well-defined and
freely chosen positions, single SPPs propagating along gold nanostructures.
Our demonstration relies on leakage-radiation microscopy of a thin
homogenous gold film and on near-field microscopy of a
nanostructure patterned in a thick gold film. The general method
offered here opens a large avenue for applications in quantum
plasmonics where an accurate positioning of quantum emitters with
respect to plasmonic conductors in a nanoscale environment is
required. For instance, it will be valuable for a locally-controlled study of the
emitter dynamics close to complex plasmonic
systems~\cite{GerardCdF,Liu,XX,YY}, e.g. to better understand the influence of the plasmonic environment on the local density of photonic states (LDOS)~\cite{GerardCdF}, to detect plasmonic non-radiative dark modes~\cite{Liu,YY}, or to map locally the photon-plasmon antibunching emission properties~\cite{GG,Greffet,Cuerto}. Other topics of a timely nature include
SPP mediated harvesting of single photons for information transfer
technology~\cite{Chang,Akimov,Fedutik,Wei,Schie}, de-multiplexing with SPPs and Bragg mirrors~\cite{multiplexer}, as well as local
addressing and interferometry~\cite{Brun,Kolesov} with single SPPs.\\

\textbf{Acknowledgments}.\\
We are grateful to F.~Treussart, J.-F.~Roch, V.~Jacques, O.~Arcizet, J.-P.~Boudou and T.~Sauvage for discussions and the nanodiamond sample, to J.-F.~Motte for the optical tip manufacturing and the gold film nanostructuration, to C.~Winkelmann and H.~Sellier for critical reading.  This work was supported by Agence Nationale de la Recherche, France, through the NAPHO and PLASTIPS projects.\\

\textbf{Supporting Information Available}. Experimental set-up and models used to simulate the near-field images. This material is available free of charge via the Internet at http://pubs.acs.org.\\
\section{Supporting Information} 

\subsection{Optical detection set-up}
The main part of the optical set up is detailed in \cite{Cuche,Cuche2}. Here we only focus on the specific part used for the plasmonics experiments described in the main article.

\begin{figure}[h]
\begin{center}
\begin{tabular}{c}
\includegraphics[width=12cm]{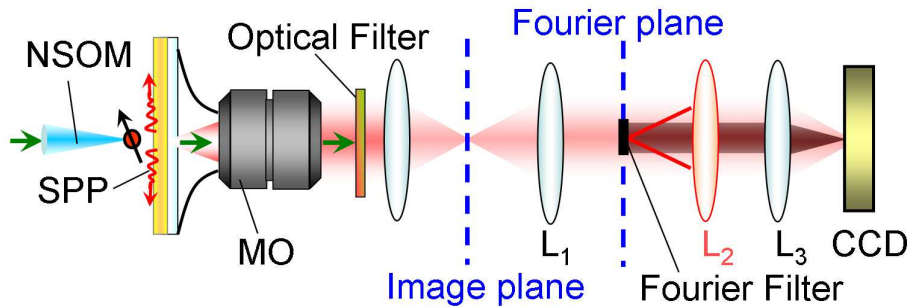}
\end{tabular}
\end{center}
\caption{Figure S1}
\end{figure}

Figure~S1 is a sketch of the NSOM microscope setup for imaging leaky
SPP waves with an oil immersion objective (MO)
[Plan Apo, NA=1.4, $100\times$ , Nikon] and a -$46^{\circ}$ Peltier cooled CCD camera [SPOT, RT-SE6, Diagnostic Instruments]. $L_1$, $L_2$, and $L_3$ are
optical lenses (achromatic doublets). A beam blocker for filtering
the contributions of allowed and forbidden lights is located in
the back focal plane of $L_1$ (which coincides with the object plane of $L_2$). When $L_2$ is removed, the SPP
propagation in the direct (spatial) space (Image Plane) is imaged.
With this lens set in, the Fourier (momentum) space (Fourier
Plane) is imaged. The principles are explained in
\cite{Drezet,Hecht,Drezet2}.

\begin{figure}[h]
\begin{center}
\begin{tabular}{c}
\includegraphics[width=12cm]{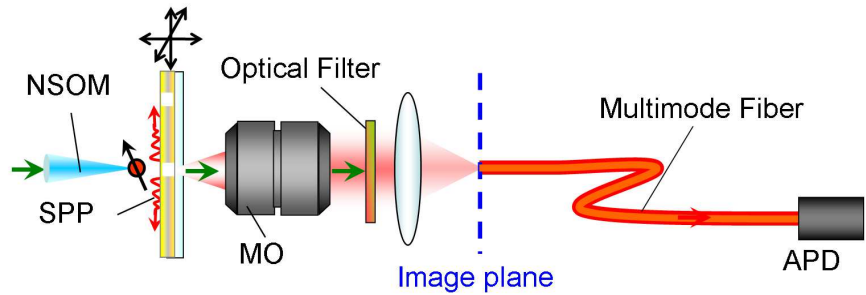}
\end{tabular}
\end{center}
\caption{Figure S2}
\end{figure}

Figure~S2 is a sketch of the NSOM microscope set-up for the scanning detection mode.  Compared to Figure~S1, a multimode optical fiber is placed in the image plane for collecting and sending the light emitted by the NV centers to an avalanche photodiode (APD) [SPCM-AQR 16, Perkin-Elmer, Canada]. The system is described in \cite{Cuche,Cuche2} and uses a dry microscope objective [CFI, Plan,  NA =0.95, $60 \times$, Nikon]. Alternatively an  Hanbury-Brown and Twiss photon correlator can be used to build up photon statistics associated with the NV emission~\cite{Cuche,Yannick,Brouri}.\\
\begin{figure}[h]
\begin{center}
\begin{tabular}{c}
\includegraphics[width=8cm]{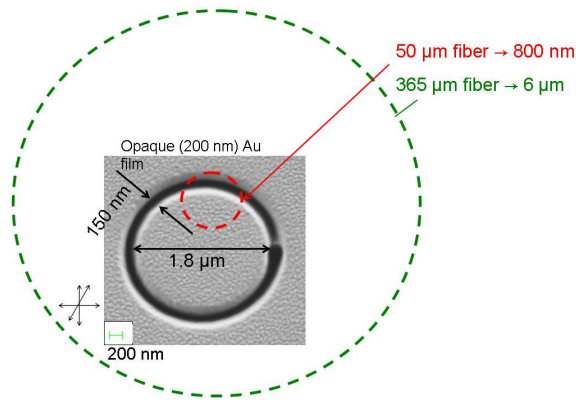}
\end{tabular}
\end{center}
\caption{Figure S3}
\end{figure}

The collection fiber conjugated with the NSOM tip defines a tolerance circular area of radius $d/2$ on the sample. The radius of this circle equals the physical radius of the fibre core $D/2$ divided by the total magnification of the microscope $M=60$: i.e. $d=D/M$. Figure S3 shows the relative size of these tolerance areas for the two multimode fibers used in the experiment. The small fiber with diameter $D=50$ $\mu$m corresponds to $d= 800$ nm. With such a fiber only the signal emitted in an area of diameter $d=800$ nm surrounding the tip can reach the detector. This means that only a short range signal emitted when the tip is in the vicinity of the ring aperture in the gold film can reach the photon detector. In contrast, the large core fiber with $D=365$ $\mu$m diameter corresponds to a tolerance area of diameter $d=6$ $\mu$m which is much larger than the plasmonic resonator size. This means than long-range signals associated with SPPs propagating from the tip to the aperture rim can reach the detector.

\subsection{Theoretical model}
The model for describing the plasmon field launched by a point-like dipole is based on the use of the dyadic green function as explained in \cite{garcia}. The main point is that for a vertical dipole the in-plane components of the electric SPP field write (up to the harmonic $e^{-i\omega t}$ dependence):
\begin{equation}
\mathbf{E}_{\textrm{SPP}}(\mathbf{r}-\mathbf{r}_0) \sim \frac{e^{iK_{\textrm{SPP}}|\mathbf{r}-\mathbf{r}_0|}}{\sqrt{(|\mathbf{r}-\mathbf{r}_0|)}}p_z\frac{(\mathbf{r}-\mathbf{r}_0)}{|\mathbf{r}-\mathbf{r}_0|}\tag{S1}
\end{equation}
whereas for a horizontal dipole they write
\begin{equation}
\mathbf{E}_{\textrm{SPP}}(\mathbf{r}-\mathbf{r}_0) \sim \frac{e^{iK_{\textrm{SPP}}|\mathbf{r}-\mathbf{r}_0|}}{\sqrt{(|\mathbf{r}-\mathbf{r}_0|)}}\left(\mathbf{p}_{||}\cdot\frac{(\mathbf{r}-\mathbf{r}_0)}{|\mathbf{r}-\mathbf{r}_0|}\right)\frac{(\mathbf{r}-\mathbf{r}_0)}{|\mathbf{r}-\mathbf{r}_0|}\tag{S2}
\end{equation}
where $\mathbf{r}:=[x,y]$ , $\mathbf{r}_0:=[x_0,y_0]$ are the in-plane coordinates of the observation point and source  point, respectively, $\mathbf{P}_{||}:=[p_x,p_y]$ are the in-plane dipole components and $K_{SPP}$ is the SPP wave vector (see eq~1 of the main article).\\   The coupling of the SPP field to the ring-like slit (with radius $R$) on the gold surface is described using the dipolar approximation. That is, we suppose that only the in-plane components of the SPP electric field launched by the point-like dipole, which is perpendicular to the rim, of the ring-like slit is coupled into radiative and propagative photons. To simulate the signal acquisition during the scan we integrate coherently the SPP field. Therefore:
\begin{equation}
I(\mathbf{r}_0)\propto \left|\int_0^{2\pi}[\mathbf{E}_{\textrm{SPP}}(\mathbf{r}(\theta)-\mathbf{r}_0)\cdot\mathbf{r}(\theta)]\mathbf{r}(\theta) d\theta \right|^2\tag{S3}
\end{equation}
where $\mathbf{r}(\theta):=[x=R\cos{(\theta)}, y=R\cos{(\theta)}]$ is the emission point on the ring. $I(\mathbf{r}_0)$ is proportional to the intensity recorded at each position of the tip during the scan.  In this first-order approximation we neglected the reflection of the SPP by the ring as well as the finite width of the slit.\\
In order to describe the coupling of the dipole field to radiative photons when the source is directly positioned above the slit the previous approximation must be modified. Here we consider that only the near-field of the dipole couples to the slit. In order to model the finite width of the slit we consider therefore the formula
\begin{equation}
I(\mathbf{r}_0)\propto \left|[\mathbf{p}_{||}\cdot\mathbf{r}_0(\theta)]\mathbf{r}_0(\theta)\right|^2 e^{-\alpha|\mathbf{r}(\theta)-\mathbf{r}_0(\theta)|^2}\tag{S4}
\end{equation}
which describes the near-field contribution with again
$$\mathbf{r}(\theta):=[x=R\cos{(\theta)}, y=R\cos{(\theta)}],$$ $$\mathbf{r}_0(\theta):=[x_0=r_0\cos{(\theta)}, y_0=r_0\cos{(\theta)}]$$
and $\alpha$ a constant.  The total transmitted signal is simulated by a linear interpolation of eqs~S3 and S4.
In the simulations of the main article we considered a vertical and an in plane dipole with components $[1,0,0]$ and $[0,1/\sqrt2,1/\sqrt2]$, but this is only an illustration since the obtained patterns are stable for a large set of dipole orientations.

\end{document}